\documentclass[aps,prd,twocolumn,superscriptaddress,showpacs]{revtex4}

\RequirePackage{lineno}

\usepackage{epsfig}
\usepackage{graphicx}
\usepackage{dcolumn}
\usepackage{bm}
\usepackage{setspace}

\begin{document}


\renewcommand{\thetable}{\Roman{table}}


\title{Observation and Mass Measurement of the Baryon $\Xi^-_b$}
\affiliation{Institute of Physics, Academia Sinica, Taipei, Taiwan 11529, Republic of China} 
\affiliation{Argonne National Laboratory, Argonne, Illinois 60439} 
\affiliation{Institut de Fisica d'Altes Energies, Universitat Autonoma de Barcelona, E-08193, Bellaterra (Barcelona), Spain} 
\affiliation{Baylor University, Waco, Texas  76798} 
\affiliation{Istituto Nazionale di Fisica Nucleare, University of Bologna, I-40127 Bologna, Italy} 
\affiliation{Brandeis University, Waltham, Massachusetts 02254} 
\affiliation{University of California, Davis, Davis, California  95616} 
\affiliation{University of California, Los Angeles, Los Angeles, California  90024} 
\affiliation{University of California, San Diego, La Jolla, California  92093} 
\affiliation{University of California, Santa Barbara, Santa Barbara, California 93106} 
\affiliation{Instituto de Fisica de Cantabria, CSIC-University of Cantabria, 39005 Santander, Spain} 
\affiliation{Carnegie Mellon University, Pittsburgh, PA  15213} 
\affiliation{Enrico Fermi Institute, University of Chicago, Chicago, Illinois 60637} 
\affiliation{Comenius University, 842 48 Bratislava, Slovakia; Institute of Experimental Physics, 040 01 Kosice, Slovakia} 
\affiliation{Joint Institute for Nuclear Research, RU-141980 Dubna, Russia} 
\affiliation{Duke University, Durham, North Carolina  27708} 
\affiliation{Fermi National Accelerator Laboratory, Batavia, Illinois 60510} 
\affiliation{University of Florida, Gainesville, Florida  32611} 
\affiliation{Laboratori Nazionali di Frascati, Istituto Nazionale di Fisica Nucleare, I-00044 Frascati, Italy} 
\affiliation{University of Geneva, CH-1211 Geneva 4, Switzerland} 
\affiliation{Glasgow University, Glasgow G12 8QQ, United Kingdom} 
\affiliation{Harvard University, Cambridge, Massachusetts 02138} 
\affiliation{Division of High Energy Physics, Department of Physics, University of Helsinki and Helsinki Institute of Physics, FIN-00014, Helsinki, Finland} 
\affiliation{University of Illinois, Urbana, Illinois 61801} 
\affiliation{The Johns Hopkins University, Baltimore, Maryland 21218} 
\affiliation{Institut f\"{u}r Experimentelle Kernphysik, Universit\"{a}t Karlsruhe, 76128 Karlsruhe, Germany} 
\affiliation{High Energy Accelerator Research Organization (KEK), Tsukuba, Ibaraki 305, Japan} 
\affiliation{Center for High Energy Physics: Kyungpook National University, Taegu 702-701, Korea; Seoul National University, Seoul 151-742, Korea; SungKyunKwan University, Suwon 440-746, Korea} 
\affiliation{Ernest Orlando Lawrence Berkeley National Laboratory, Berkeley, California 94720} 
\affiliation{University of Liverpool, Liverpool L69 7ZE, United Kingdom} 
\affiliation{University College London, London WC1E 6BT, United Kingdom} 
\affiliation{Centro de Investigaciones Energeticas Medioambientales y Tecnologicas, E-28040 Madrid, Spain} 
\affiliation{Massachusetts Institute of Technology, Cambridge, Massachusetts  02139} 
\affiliation{Institute of Particle Physics: McGill University, Montr\'{e}al, Canada H3A~2T8; and University of Toronto, Toronto, Canada M5S~1A7} 
\affiliation{University of Michigan, Ann Arbor, Michigan 48109} 
\affiliation{Michigan State University, East Lansing, Michigan  48824} 
\affiliation{University of New Mexico, Albuquerque, New Mexico 87131} 
\affiliation{Northwestern University, Evanston, Illinois  60208} 
\affiliation{The Ohio State University, Columbus, Ohio  43210} 
\affiliation{Okayama University, Okayama 700-8530, Japan} 
\affiliation{Osaka City University, Osaka 588, Japan} 
\affiliation{University of Oxford, Oxford OX1 3RH, United Kingdom} 
\affiliation{University of Padova, Istituto Nazionale di Fisica Nucleare, Sezione di Padova-Trento, I-35131 Padova, Italy} 
\affiliation{LPNHE, Universite Pierre et Marie Curie/IN2P3-CNRS, UMR7585, Paris, F-75252 France} 
\affiliation{University of Pennsylvania, Philadelphia, Pennsylvania 19104} 
\affiliation{Istituto Nazionale di Fisica Nucleare Pisa, Universities of Pisa, Siena and Scuola Normale Superiore, I-56127 Pisa, Italy} 
\affiliation{University of Pittsburgh, Pittsburgh, Pennsylvania 15260} 
\affiliation{Purdue University, West Lafayette, Indiana 47907} 
\affiliation{University of Rochester, Rochester, New York 14627} 
\affiliation{The Rockefeller University, New York, New York 10021} 
\affiliation{Istituto Nazionale di Fisica Nucleare, Sezione di Roma 1, University of Rome ``La Sapienza," I-00185 Roma, Italy} 
\affiliation{Rutgers University, Piscataway, New Jersey 08855} 
\affiliation{Texas A\&M University, College Station, Texas 77843} 
\affiliation{Istituto Nazionale di Fisica Nucleare, University of Trieste/\ Udine, Italy} 
\affiliation{University of Tsukuba, Tsukuba, Ibaraki 305, Japan} 
\affiliation{Tufts University, Medford, Massachusetts 02155} 
\affiliation{Waseda University, Tokyo 169, Japan} 
\affiliation{Wayne State University, Detroit, Michigan  48201} 
\affiliation{University of Wisconsin, Madison, Wisconsin 53706} 
\affiliation{Yale University, New Haven, Connecticut 06520} 
\author{T.~Aaltonen}
\affiliation{Division of High Energy Physics, Department of Physics, University of Helsinki and Helsinki Institute of Physics, FIN-00014, Helsinki, Finland}
\author{A.~Abulencia}
\affiliation{University of Illinois, Urbana, Illinois 61801}
\author{J.~Adelman}
\affiliation{Enrico Fermi Institute, University of Chicago, Chicago, Illinois 60637}
\author{T.~Affolder}
\affiliation{University of California, Santa Barbara, Santa Barbara, California 93106}
\author{T.~Akimoto}
\affiliation{University of Tsukuba, Tsukuba, Ibaraki 305, Japan}
\author{M.G.~Albrow}
\affiliation{Fermi National Accelerator Laboratory, Batavia, Illinois 60510}
\author{S.~Amerio}
\affiliation{University of Padova, Istituto Nazionale di Fisica Nucleare, Sezione di Padova-Trento, I-35131 Padova, Italy}
\author{D.~Amidei}
\affiliation{University of Michigan, Ann Arbor, Michigan 48109}
\author{A.~Anastassov}
\affiliation{Rutgers University, Piscataway, New Jersey 08855}
\author{K.~Anikeev}
\affiliation{Fermi National Accelerator Laboratory, Batavia, Illinois 60510}
% ...
\collaboration{CDF Collaboration\footnote{With visitors from $^a$University of Athens, 15784 Athens, Greece, 
$^b$University of Bristol, Bristol BS8 1TL, United Kingdom, 
$^c$University Libre de Bruxelles, B-1050 Brussels, Belgium, 
$^d$Cornell University, Ithaca, NY  14853, 
$^e$University of Cyprus, Nicosia CY-1678, Cyprus, 
$^f$University College Dublin, Dublin 4, Ireland, 
$^g$University of Edinburgh, Edinburgh EH9 3JZ, United Kingdom, 
$^h$University of Heidelberg, D-69120 Heidelberg, Germany, 
$^i$Universidad Iberoamericana, Mexico D.F., Mexico, 
$^j$University of Manchester, Manchester M13 9PL, England, 
$^k$Nagasaki Institute of Applied Science, Nagasaki, Japan, 
$^l$University de Oviedo, E-33007 Oviedo, Spain, 
$^m$University of London, Queen Mary College, London, E1 4NS, England, 
$^n$University of California Santa Cruz, Santa Cruz, CA  95064, 
$^o$Texas Tech University, Lubbock, TX  79409, 
$^p$University of California, Irvine, Irvine, CA  92697, 
$^q$IFIC(CSIC-Universitat de Valencia), 46071 Valencia, Spain, 
}}
\noaffiliation

\pacs{13.30.Eg, 13.60.Rj, 14.20.Mr}

\begin{abstract}
We report the observation and measurement of 
the mass of the bottom, strange baryon
$\Xi^-_b$ through the decay chain $\Xi^-_b \rightarrow J/\psi \, \Xi^-$,
where $J/\psi \rightarrow \mu^+\, \mu^-$, $\Xi^- \rightarrow \Lambda \, \pi^-$,
and $\Lambda \rightarrow p \, \pi^-$.
A signal is observed whose probability of arising 
from a background fluctuation is $6.6 \times 10^{-15}$, or 
7.7 Gaussian standard deviations.
The $\Xi^-_b$ mass is measured to be 
$5792.9\pm 2.5$ (stat.) $\pm 1.7$ (syst.) MeV/$c^2$.
\end{abstract}
\maketitle

Since its inception, the quark model has had great success in describing
the spectroscopy of hadrons.  
The quark model has been successful for the $B$ mesons, where all of the 
ground states have been observed \cite{PDG}.
The spectroscopy of $c$ baryons also agrees well with the quark model, 
and a  rich
spectrum of baryons containing $b$ quarks is predicted \cite{Jenkins}.
However, direct observation of $b$ baryons 
has been limited to a single state, the
$\Lambda_b$ (quark content $udb$) \cite {PDG}, until recently.  
Evidence for $b$ baryons that also contain
a strange quark was shown from LEP \cite{LEP} through 
partial reconstruction of decays containing electrons and muons.
Recent results from the 
Tevatron on the $\Sigma_b$ states (quark content $uub, ddb$) 
\cite{CDF_Sigma_b}
and $\Xi^-_b$ (quark content $dsb$) \cite{DZero}
are beginning to subject the $b$ baryons to closer examination.


In this Letter, we report the observation of
a heavy baryon 
and  measurement of its mass.
The decay properties of this state are consistent with the weak decay
of a $b$ baryon, and we interpret our result as the observation
of the $\Xi_b^-$ baryon.
This observation is made in
$p\overline{p}$ collisions 
at a center of mass energy of 1.96 TeV using the
Collider Detector at Fermilab (CDF II), through the decay chain 
$\Xi^-_{b} \rightarrow$ $J/\psi \, \Xi^-$, where
$J/\psi \rightarrow \mu^+\mu^-$, $\Xi^- \rightarrow \Lambda \, \pi^-$,
and $\Lambda \rightarrow p \, \pi^-$.
Charge conjugate modes are included implicitly.
This measurement is based on 
a data sample with an integrated luminosity of
1.9 fb$^{-1}$.  


The CDF II detector has been described in detail elsewhere 
\cite{CDF_detector}.  
This analysis primarily relies upon the tracking and 
muon identification systems.  
The tracking system consists of a
silicon microstrip detector 
and an open-cell drift chamber (COT)
that operate inside a solenoid with a 1.4 T magnetic field.
 Muon candidates from the decay $J/\psi \rightarrow \mu^+ \mu^-$
 are  identified by  two sets of drift  chambers located radially
 outside  the   electromagnetic and   hadronic  calorimeters.
The central muon chambers
cover the pseudorapidity region $ |\eta| <  0.6$ \cite{pt_def}, 
and  are sensitive to muons with 
transverse momentum $p_T > 1.4 $ GeV/$c$.
A second muon system covers the region $  0.6 < |\eta| <  1.0$ and
detects  muons having $p_T > 2.0 $ GeV/$c$.  
Muon triggering and identification
are based on matching tracks measured in the muon system to COT tracks.


The analysis of the data begins with a selection of well-measured
$J/\psi \rightarrow \mu^+ \mu^-$ candidates.
The trigger requirements are confirmed by selecting
 events that contain two
 oppositely  charged  muon candidates, each with matching 
 COT and muon chamber tracks.
 We also require that both muon tracks have associated 
measurements in at least three layers of the silicon detector and
a two-track invariant  mass within 80 MeV/$c^2$ of the
world-average  $J/\psi$ mass \cite{PDG}. 
This data sample provides approximately 15 million events containing $J/\psi$
candidates, measured with an average mass resolution of 20 MeV/$c^2$.
 


The reconstruction of $\Xi^-$
candidates uses all additional tracks 
found in each selected $J/\psi$ event.  
Pairs of oppositely charged tracks are used to identify
$\Lambda$ decay candidates.
The proton (pion) mass is assigned 
to the track with the higher (lower) momentum.
This mass assignment is 
always correct for $\Lambda \rightarrow p \, \pi^-$ candidates 
used in this analysis
because of the
kinematics of $\Lambda$ decay and the lower limit of 
$\approx$200 MeV/$c$ in the transverse momentum
acceptance of the tracking system. 
The $\Lambda$ mass is measured with a resolution of 2.5 MeV/$c^2$.
All intersecting pairs of tracks with an invariant mass
within 10 MeV/$c^2$  of the world average $\Lambda$ mass \cite{PDG}
have their
track parameters recalculated according to a fit where
the momenta of the two tracks are constrained 
to the $\Lambda$ mass.
The decay vertex is used to calculate the 
$\Lambda$ displacement from the beam-line in the direction of the
track pair's   transverse  momentum.  
The background due to tracks originating from the primary vertex
is reduced by requiring this displacement to exceed 1.0 cm.  
For candidates that satisfy these  requirements,
the remaining tracks are assigned the pion mass, and
$\Lambda \, \pi^-$ combinations 
are identified that are consistent with the decay process
$\Xi^- \rightarrow \Lambda \, \pi^-$.  In order to obtain the best possible 
$\Lambda \, \pi^-$ mass resolution, the reconstruction uses
a fit on the three tracks that simultaneously 
constrains the $\Lambda$
decay products to the $\Lambda$ mass, and the
$\Lambda$ trajectory to intersect with the helix of the $\Xi^-$ decay pion.
For all $\Xi^-$ candidates, the reconstructed decay
position of the $\Lambda$ candidate is required to be radially 
displaced at least 1.0 cm with respect to
the reconstructed decay vertex of the $\Xi^-$ candidate.

\begin{figure}[p]
\psfig{figure=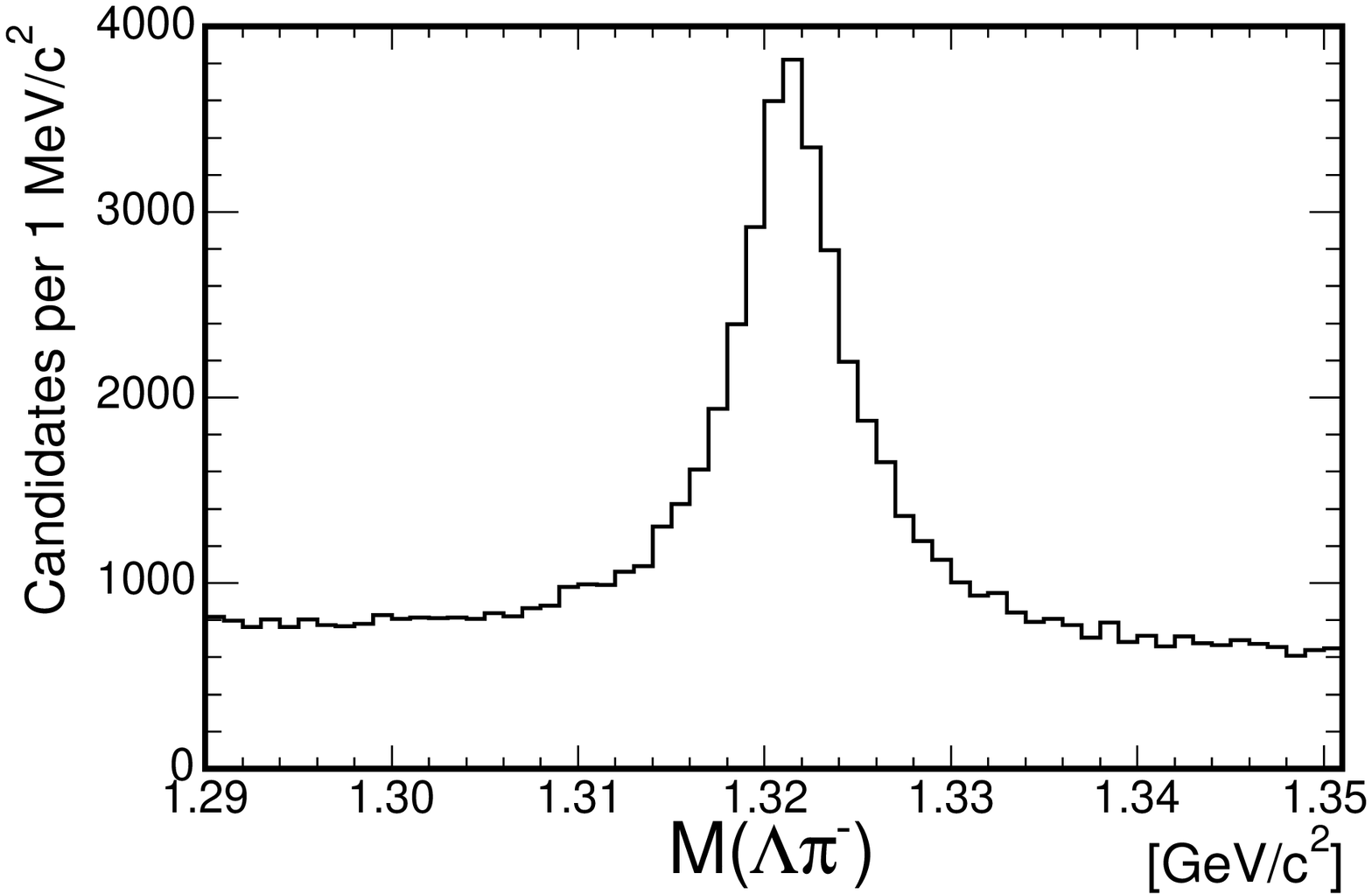, width=3.0in}
\caption{The invariant mass distribution of $\Lambda \, \pi^-$ combinations 
having an associated track in the silicon detector in events containing $J/\psi$ candidates.
\label{fig:xi} }
\end{figure}

The majority of $\Xi^-$ candidates have $p_T >$ 1.5 GeV/$c$.  
This, along with the long lifetime of the $\Xi^-$ ($c\tau$ = 4.9 cm) \cite{PDG},
results in a significant fraction of the $\Xi^-$ candidates having
decay vertices located several centimeters radially outward from the beam-line.
Therefore, we are able to
refine the $\Xi^-$ reconstruction by making use of the 
improved determination of the trajectory that can be obtained by tracking 
the $\Xi^-$ in
the silicon detector.  The $\Xi^-$ candidates have an additional
fit performed on the three tracks  that simultaneously 
constrains
both the $\Lambda$ and $\Xi^-$ masses of the appropriate track combinations,
and provides the best possible estimate of the $\Xi^-$ momentum and decay
position.
The result of this fit is used to define a helix that
serves as the seed for an algorithm that searches for silicon detector
hits associated with 
the $\Xi^-$ track.  We retain for further analysis all $\Xi^-$ 
candidates with measurements in at least two layers of the 
silicon detector.
This technique provides excellent impact parameter resolution for the
$\Xi^-$ track (average of 60 $\mu$m), 
and has been used previously \cite{CDF_PentaQ}.
The $\Lambda \, \pi^-$ invariant mass spectrum of all combinations
that satisfy these requirements is shown in Fig. \ref{fig:xi}.
Approximately 23,500 $\Xi^-$ candidates above the combinatorial
background are identified.  
The $\Xi^-_b$ search includes 
the subset of these combinations with an invariant
mass within 10 MeV/$c^2$ of the world average $\Xi^-$ mass \cite{PDG}.

The mass resolution for the
$J/\psi \, \Xi^-$ final state is
studied with a Monte Carlo simulation that generates $b$ quarks
according to a next-to-leading-order calculation \cite{NDE},
and produces  
$\Xi^-_{b}$ events by simulating 
$b$ quark fragmentation \cite{Peterson}.
The decay $\Xi^-_{b} \rightarrow J/\psi \, \Xi^-$ is simulated with 
EvtGen \cite{EvtGen}.
The generated events are 
used as input to the detector and trigger simulations based on a 
GEANT3 description \cite{Geant}  
and processed through 
the same reconstruction and
analysis algorithms used for the data.
Analysis of the simulated $\Xi^-_{b}$ events
shows that a 10\% improvement in mass resolution can be obtained
if the momenta of the $\Xi^-$ decay products are allowed to vary
in the fit of the $\Xi^-_{b}$ candidate, rather than simply using
the $\Xi^-$ track.
Consequently, a procedure that simultaneously fits the five tracks 
of the final state, constrains the
three vertices of the decay chain to the 
appropriate topology, and constrains 
the masses of the $J/\psi$, $\Xi^-$, and $\Lambda$ to their world 
average masses, is used to provide the best estimate of the
$J/\psi \, \Xi^-$ mass.  
The average $\Xi^-_{b}$ mass resolution obtained from simulated events
is found to be approximately $15$ MeV/$c^2$.  In particular, 
we note that the $J/\psi \, \Xi^-$ invariant mass resolution is
comparable to
the mass resolution obtained with the CDF II detector for other $B$
hadrons with a $J/\psi$ in the final state \cite{CDF_Bmass}.

The selection used to isolate the 
$\Xi^-_{b} \rightarrow$ $J/\psi \, \Xi^-$ decay process is guided
by the properties of other $B$ hadrons 
that include a $J/\psi$ in the final state.  
The important ones include the lifetime of the ground-state
$B$ hadrons and the energy available in the decay.
The $B^\pm, B^0$, and $B_s$ mesons and
$\Lambda_b$ baryon, all  appear to have lifetimes dominated by 
the weak decay of the $b$ quark.  We expect the same to hold
true for the $\Xi^-_{b}$ and for its lifetime to be comparable to  
these states.  In addition,
these particles all decay to final states $J/\psi \, X$, where $X$ is a single
hadron.  In these decays the momentum carried by the decay products
in the rest frame of the $B$ hadron
falls in the fairly narrow range of $1570 - 1744$ MeV/$c^2$
for decays where only the lightest decay products are considered.
We expect the energy released in the decay of the
$\Xi^-_{b}$ to be comparable.  
This expectation is also consistent with the range of theoretical 
predictions ($5788-5812$ MeV/$c^2$)
for the $\Xi^-_b$ mass \cite{Jenkins}.
Consequently, we have chosen
a $\Xi^-_{b}$ search strategy that will provide an optimal sensitivity
for a final state that shares similar properties with these well-established
$B$ hadrons.


We have developed the event
selection by studying $B^\pm \rightarrow J/\psi \, K^\pm$
decays.  This final state is identified by assigning the $K^\pm$ mass to
all tracks not used in the $J/\psi$ reconstruction.
Each three-track combination must satisfy 
a fit where the tracks are required to originate from a common
vertex and the invariant mass of the muon pair is constrained to the
world average $J/\psi$ mass \cite{PDG}.
Approximately 30,000 $B^\pm$ candidates are identified in this sample.
Several characteristics of the final state are used as selection 
requirements to obtain a $B^\pm$ signal
with very little background.
Minimum transverse momentum requirements 
on the $K^\pm$ and $B^\pm$  candidates are used to suppress
backgrounds from the event that are not related to the $B^\pm$ 
decay.
The trajectory of the $K^\pm$  is required to originate from the
$B^\pm$ decay vertex by placing a requirement on its impact parameter
$d_{SV}(K)$ and associated uncertainty $\sigma_{d_{SV}}(K)$
with respect to the vertex found in the 
$J/\psi$ fit.
Similar impact parameter quantities $d_{PV}(K)$ and $\sigma_{d_{PV}}(K)$
measured with respect to the primary vertex are used to remove tracks
that originate from the prompt background.
Reasonable vertex quality is assured by placing a minimum value on the
accepted probability $P(\chi^2)$ of the mass- and vertex-constrained 
fit used to obtain the $B^\pm$ candidate.
We suppress the promptly-produced combinatorial background by rejecting 
candidates with low proper decay time, 
$t \equiv \vec{L_T} \cdot \vec{p_T}(B) \frac{M(B)}{|p_T(B)|^2}$, 
where $M(B)$ is the mass of the $B^\pm$ candidate,
$\vec{p_T}(B)$ is the transverse momentum of the $B^\pm$ candidate, and
$\vec{L_T}$ is the transverse displacement 
of the $B^\pm$ decay vertex from the beam-line. 
A requirement on proper decay time uncertainty $\sigma_{t}$
removes poorly-reconstructed combinations.
We also reject combinations that are inconsistent with having
originated from the beam-line by requiring a small magnitude of
the impact of the $B^\pm$ candidate,
$\vec{d}_{PV}(B) \equiv \vec{L_T} \times \vec{p_T}(B)/|p_T(B)|$, and a small
angle $\beta$ between  $\vec{L_T}$ and $\vec{p_T}(B)$.

This analysis uses a two-step selection procedure.
Final selection criteria are listed in 
Table \ref{table:selection}.  We first impose the ``standard" selection
requirements listed there and retain all $J/\psi \, K^\pm$ 
combinations that satisfy them.  
Any combination that fails only one of the ``standard" selection
requirements
is also allowed into the final sample if it satisfies both of the 
$p_T$ requirements of the ``high-$p_T$" selection requirements and
fails no more than one of the other requirements in this set.  
The combination of ``standard" and ``high-$p_T$" requirements
reduce the background in the 
$B^\pm \rightarrow J/\psi \, K^\pm$ sample
to approximately 400 combinations, while retaining a signal of 
16,000 $B^\pm$ candidates.

\begin{figure}[p]
\psfig{figure=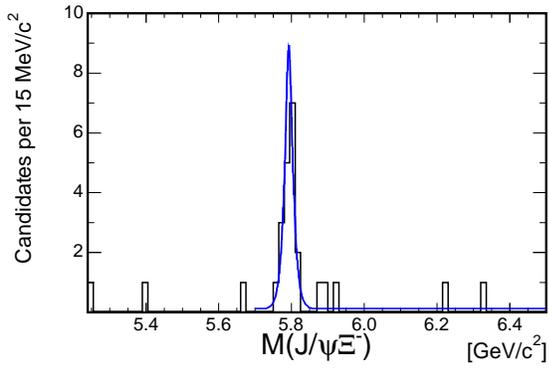,width=3.0in}
\caption{The $J/\psi \, \Xi^-$ invariant mass distribution for 
combinations that satisfy the selection requirements.  
The projection of the fit function is overlaid on the data.
\label{fig:JpsiXi} }
\end{figure}

As is done for the reconstruction of the $B^\pm$, the treatment of
the $J/\psi \, \Xi^-$ candidates requires a mass- 
and vertex-constrained fit on the muon
candidates and the $\Xi^-$ track.  The selection criteria in 
Table \ref{table:selection} are applied
to the $J/\psi \, \Xi^-$ sample
where we simply exchange $K^-$ for $\Xi^-$ and $B^\pm$ for $\Xi^-_b$  
where appropriate.  
Combinations that satisfy these requirements form the final set of
$\Xi^-_b$ candidates.
The invariant mass of each candidate is obtained with the 
full five-track fit, and the resulting $J/\psi \, \Xi^-$ mass distribution
is shown in Fig. \ref{fig:JpsiXi}.

\begin{table}[ht]
\begin{center}
\caption{Selection variables and requirements for the ``standard" selection
and ``high-$p_T$" selection as
described in the text. Here ``$K$'' refers to the third track combined with
the $J/\psi$ and is a $K^\pm$ or $\Xi^-$ candidate for the $B^\pm$ or 
$\Xi^-_b$ candidates, respectively.
Similarly, ``$B$" refers to either $B^\pm$ or $\Xi^-_b$, as is appropriate.
\label{table:selection}}
\begin{tabular}{ccc}
\hline \hline
Selection variable & Standard & High-$p_T$ \\
\hline
 $p_T(K)$ & $> 1.7$ GeV/$c$ & $> 2.5$ GeV/$c$ \\
 $p_T(B)$ & $> 5$ GeV/$c$ & $> 6$ GeV/$c$ \\
\hline
 $|d_{SV}(K)|$  & $< 100~\mu$m & $< 80~\mu$m \\
 $|d_{PV}(K)|/ \sigma_{d_{PV}(K)}$ & $> 2.5 $ & $> 3  $\\
 $P(\chi^{2})$ & $> 0.1\% $ & $> 1\% $ \\
 $|d_{PV}(B)|$  & $< 75~\mu$m & $< 60~\mu$m \\
 $ct$ & $> 80~\mu$m & $> 100~\mu$m \\
 $c\sigma_{t}$ & $< 30~\mu$m & $< 25~\mu$m \\
 $\beta$ & $< 0.4$ radians & $< 0.3$ radians \\
\hline \hline
\end{tabular}
\end{center}
\end{table}


The mass  resolution estimate for the $\Xi^-_{b}$ 
implies that more than $95\%$ of  a $\Xi^-_{b}$
signal will occupy an invariant mass bin with a width of  75 MeV/$c^2$.
The data shown in Fig. \ref{fig:JpsiXi} contain 18 candidates in the
75 MeV/$c^2$ range of $5750-5825$ MeV/$c^2$.  
We model the combinatorial background by considering candidates 
in the range $5700-6500$ MeV/$c^2$; the data yield 23 
candidates in this range.  
No events contribute multiple candidates.
The upper limit of this range is chosen arbitrarily, and 
has no impact on the result.  The lower limit is chosen
to avoid partially reconstructed 
$\Xi^{-,0}_b \rightarrow J/\psi \, \Xi^- \, X$
decays, where $X$ represents additional undetected particles.
We assume that the mass distribution of
the combinatorial background is uniform and that the occupancy 
due to background combinations in any
particular 75 MeV/$c^2$ mass bin 
within the 800 MeV/$c^2$ search range 
can be described by a binomial distribution, with a single
event probability given by
the ratio of the two mass ranges used.
The probability that 
the number of candidates observed in the $5750-5825$ MeV/$c^2$ mass range
is due to a background fluctuation is
estimated as the binomial probability of 18 or more
events from a sample of 23 total occurences
and a single event probability of $75/800$.
This probability 
is $6.6 \times 10^{-15}$, equivalent to a $7.7 \sigma$ 
variation from a Gaussian distribution.  Consequently, we 
interpret the data distribution shown in Fig. \ref{fig:JpsiXi} to be the 
observation of a resonance, with a width consistent with the 
detector resolution.
Comparable distributions of $J/\psi \, \Lambda \, \pi^+$ and
$J/\psi \, \Lambda \, \pi^-$ where the $\Lambda \, \pi^-$ does not form a 
$\Xi^-$, yield no significant enhancement at any mass within the 
range of this analysis.

The masses and their uncertainties
obtained from the five-track final state fit method are 
used in an unbinned likelihood fit to measure
the $\Xi_b^-$ mass.
The negative log-likelihood function that is minimized has the form
\begin{equation}
{\cal L} = -2\sum_{i=1}^{N} \ln \left[ 
fG(m_i,m_0,s_m\sigma^m_i) + (1-f) {\cal C}\right ],
\label{eq:unbinned}
\end{equation}
where $m_i$ is the mass obtained for a single candidate, 
$\sigma^m_i$ is the uncertainty on that mass as estimated from the
track parameters for the candidate,
$G(m_i,m_0,s_m\sigma^m_i)$ is a Gaussian distribution with average
$m_0$ and characteristic width $s_m\sigma^m_i$, and 
${\cal C}$ is a constant background term.  
The quantities
obtained from the fitting procedure include $f$, the fraction of the events
found in the signal, $m_0$, the best average mass, and $s_m$, a scale
factor on the mass uncertainty 
which accounts for a possible shift in the mass uncertainty.
This procedure yields a best estimate for the 
$\Xi_b^-$ mass of $5792.9\pm2.5$ MeV/$c^2$.  The uncertainty scale factor
is determined to be $2.0\pm0.4$, which is consistent with the value
of $1.6\pm0.2$ obtained with simulated $\Xi^-_b$ candidates.  
The signal fraction  is calculated to be 
$0.76\pm0.09$, giving a yield of $17.5\pm4.3$ $\Xi^-_b$ candidates in this 
data sample.
The projection of the fit function is superimposed on Fig. \ref{fig:JpsiXi}.


Several systematic effects have been considered for their impact on the
$\Xi^-_b$ mass measurement.  The overall momentum scale of the 
tracking system is established by calibrating with the well-measured 
$J/\psi$, $\psi(2S)$, and $\Upsilon$ states \cite{CDF_Bmass}.
The momentum scale uncertainty contributes a systematic uncertainty of 
$\pm0.4$ MeV/$c^2$ on the $\Xi^-_b$ mass measurement.  
Alignment and material distribution of the tracking system contributes
an additional $\pm0.6$ MeV/$c^2$ of uncertainty.
The method
requires knowledge of the mass of the final state $\Xi^-$, 
a quantity that
is known to $\pm0.13$ MeV/$c^2$ \cite{PDG}.  
The mass uncertainty of the $J/\psi$ is included in the 
momentum scale calibration, and the mass uncertainty of the $\Lambda$
($\pm0.006$ MeV/$c^2$) is negligible for this analysis.
Finally, the largest systematic variation seen on the mass measurement
occurs when alternative fitting models are used for the mass calculation.
The signal distribution has been fit with a 
single Gaussian where its parameters are allowed to vary in the fit.
A double Gaussian is also used, where the widths are fixed to
values obtained in simulation, and only the average mass is allowed
to vary.
These different models for the probability distribution function of the 
signal create variations of $\pm1.5$ MeV/$c^2$ on the mass result.  
The individual systematic uncertainties are combined in quadrature to obtain
an overall systematic uncertainty of $\pm1.7$ MeV/$c^2$ on the
$\Xi^-_b$ mass measurement.


In conclusion, we use the CDF II detector at the Tevatron to
observe the $\Xi^-_b$ in $p\overline{p}$ collisions.
A signal with $17.5\pm4.3$ $\Xi^-_b$ candidates and 
a significance of $7.7\sigma$ is seen in the decay channel
$\Xi^-_b \rightarrow J/\psi \, \Xi^-$.  The mass of this baryon
is measured to be $5792.9\pm2.5$ (stat.) $\pm1.7$ (syst.) MeV/$c^2$,
which is consistent with theoretical expectations 
\cite{Jenkins}.
The mass measurement presented is also consistent with the only 
other direct
observation of this state \cite {DZero}, and represents a 
significant improvement in precision.

We thank the Fermilab staff and the technical staffs of the participating
 institutions for their vital contributions. This work was supported by the
 U.S. Department of Energy and National Science Foundation; the Italian
 Istituto Nazionale di Fisica Nucleare; the Ministry of Education, Culture,
 Sports, Science and Technology of Japan; the Natural Sciences and Engineering
 Research Council of Canada; the National Science Council of the Republic of
 China; the Swiss National Science Foundation; the A.P. Sloan Foundation; the
 Bundesministerium f\"ur Bildung und Forschung, Germany; the Korean Science
 and Engineering Foundation and the Korean Research Foundation; the Science
 and Technology Facilities Council and the Royal Society, UK; the Institut
 National de Physique Nucleaire et Physique des Particules/CNRS; the Russian
 Foundation for Basic Research; the Comisi\'on Interministerial de Ciencia
 y Tecnolog\'{\i}a, Spain; the European Community's Human Potential Programme;
 the Slovak R\&D Agency; and the Academy of Finland.

\begin{thebibliography}{5}
\bibitem{pt_def}The pseudorapidity $\eta$ is defined as
$\eta \equiv -\ln ( \tan( \theta /2) )$, where $\theta$ 
is the angle between the particle
momentum and the proton beam direction.
The transverse momentum $p_T$ is defined as the component of the particle
momentum perpendicular to the proton  beam direction.
\end{thebibliography}
\end{document}